# Parylene Based Memristive Devices with Multilevel Resistive Switching for Neuromorphic Applications

Anton A. Minnekhanov[1,*], Andrey V. Emelyanov[1,2], Dmitry A. Lapkin[1,†], Kristina E. Nikiruy[1,2], Boris S. Shvetsov[3], Alexander A. Nesmelov[1], Vladimir V. Rylkov[1,4], Vyacheslav A. Demin[1,2], Victor V. Erokhin[1,5]

[1]National Research Center "Kurchatov Institute", 123182 Moscow, Russia.

[2]Moscow Institute of Physics and Technology, 141700 Dolgoprudny, Moscow Region, Russia.

[3]Lomonosov Moscow State University, 119991 Moscow, Russia.

[4]Kotel'nikov Institute of Radio Engineering and Electronics RAS, 141190 Fryazino, Moscow Region, Russia.

[5]CNR-IMEM (National Research Council, Institute of Materials for Electronics and Magnetism), 43124 Parma, Italy

[†]Present address: Deutsches Elektronen-Synchrotron DESY, 22607 Hamburg, Germany.

*E-mail: minnekhanov@physics.msu.ru, minnekhanov_aa@nrcki.ru

**Abstract**

In this paper, the resistive switching and neuromorphic behavior of memristive devices based on parylene, a polymer both low-cost and safe for the human body, is comprehensively studied. The Metal/Parylene/ITO sandwich structures were prepared by means of the standard gas phase surface polymerization method with different top active metal electrodes (Ag, Al, Cu or Ti of ~ 500 nm thickness). These organic memristive devices exhibit excellent performance: low switching voltage (down to 1 V), large OFF/ON resistance ratio (~ $10^3$), retention ($\geq 10^4$ s) and high multilevel resistance switching (at least 16 stable resistive states in the case of Cu electrodes). We have experimentally shown that parylene-based memristive elements can be trained by a biologically inspired spike-timing-dependent plasticity (STDP) mechanism. The obtained results have been used to implement a simple neuromorphic network model of classical conditioning. The described advantages allow considering parylene-based organic memristors as prospective devices for hardware realization of spiking artificial neuron networks capable of supervised and unsupervised learning and suitable for biomedical applications.

**Introduction**

Memristive devices are of great research interest nowadays owing to a number of their attractive properties such as low energy consumption and voltage operation, high write/read rate, multilevel resistive switching (RS) and ability to store intermediate states as well as manufacturability due to their simple two-terminal structure, and low cost of fabrication [1–4]. Among various applications of memristive devices, the most promising are resistive random-access memory (RRAM) fabrication [1], computing in memory [2], and neuromorphic computing [3–8]. The multilevel character of RS, or the possibility of "analog" resistance variation in the window between the low- and high-resistance states ($R_{on}$, $R_{off}$), is one of the most important properties of memristors for emulating synapses (key elements of biological neural networks that couple neurons with variable weight functions) in the development of neuromorphic computing systems like the human brain (in terms of pattern and speech recognition, learning ability, and other cognitive tasks) [2, 3].

It is also important that there are organic memristive structures tnat could be fabricated on flexible biocompatible substrates, used for neuroprosthetics and so-called "wearable" electronic devices [5, 9, 10]. Stable operation of flexible memristive devices was shown at rather small bending radii (down to 9 mm) and longitudinal twisting up to 30 degrees [11, 12]. Moreover, resistance in this case can be tuned in the window between $R_{on}$ and $R_{off}$ according to the rules similar to those in biological neural networks [5], in particular, the so-called "spike-timing-dependent plasticity" (STDP) first implemented for inorganic memristive structures [13].

The memristive properties were found in structures based on inorganic oxides ($TiO_x$, $HfO_x$, $SiO_x$, $TaO_x$, etc.) [14–18], organic (polyaniline, polythiophene) [19–21] and nanocomposite [22] materials. Most memristive devices operate through the electromigration of oxygen vacancies in oxides and formation (rupture) of conductive filaments (valence change memristors) or metal bridge growth (destruction) by means of cation motion in a dielectric matrix (electrochemical metallization or ECM memristors) [1]. In the last case an electrochemically active metal, for example Ag, is used as one of the electrodes of the memristive structure of metal-insulator-metal (MIM) type. Under applied positive voltage, $Ag^+$ cations can migrate to the cathode, where they are reduced and form metal bridges [1, 5, 23–30]. This mechanism can also be realized with Cu [30–32] and Al [33–35] active electrodes. Moreover, the conductive filaments can consist of carbon formed by local pyrolysis of organic material [36, 37].

One of the most promising memristive structures for "wearable" applications at the present moment are connected to MIM structures based on polymeric layers of parylene (poly-para-xylylene, or PPX) due to the simple and cheap production of this polymer, its transparency and the possibility of the fabrication on flexible substrates [35, 38]. Moreover, parylene is an FDA-approved material and could be used in biomedical applications since it is completely safe for the human body, which cannot be said about most of the other organic materials [35, 39, 40]. Parylene-based materials have found the widest applications in electronics and electrical engineering, and, especially, in radio-electronic equipment production. Parylene is used as protective coating and an insulating layer in integrated circuits and thin-film transistors, microelectromechanical systems, lasers, waveguides and photodiodes [41–43].

Currently, parylene-based memristive structures have shown a fairly wide window of RS ($R_{off}/R_{on} \sim 10^4$) and reasonable retention [35, 38]. However, the possibility of their "analog" resistance variation, or plasticity (an analog of synaptic plasticity in biological systems), as well as their implementation in neuromorphic systems, has not been reported yet. Therefore, the main goals of this study are: 1) to develop a technology for fabrication of memristive structures based on parylene layers with multilevel resistance switching (high plasticity); 2) to study their memristive properties, as well as the STDP learning ability; 3) to demonstrate the ability of parylene-based memristors to implement synapse-like elements in spiking neuromorphic networks with STDP learning.

**Materials and Methods**

We have studied memristive elements of Metal/Parylene/ITO structure (further M/PPX/ITO structure). The parylene layers (~ 100 nm) were deposited on a commercially purchased ITO coated glass substrate (bottom electrode) by the gas phase surface polymerization method using SCS Labcoater PDS 2010 vacuum deposition system. PDS 2010 transforms Parylene dimer (2,2-para-cyclophane or its derivatives) to a gaseous monomer; the material polymerizes onto the substrate upon deposition at room temperature. There is no intermediate liquid phase or separate cure cycle. At the vacuum levels employed, all sides of the substrate were uniformly impinged on by the gaseous monomer, resulting in a truly conformal coating.

The top metal electrodes were Ag, Al, Cu or Ti layers (~ 500 nm) obtained by thermal evaporation or ion-beam sputtering through a shadow mask. The sizes of the top electrodes were 0.2×0.5 mm$^2$, and about 150 devices (per one substrate) were fabricated for each kind of electrode. The metals listed above were selected due to their widespread use in electronic engineering including the manufacture of memristive devices.

The structural investigations were carried out with the transmission electron microscope (TEM) Titan 80-300 (FEI, USA) in TEM and STEM modes. An energy dispersive X-RAY analyzer (EDX, USA) was used to reveal the chemical composition. The cross-sectional preparation of the memristive samples was done by the focused ion beam (FIB) method on Helios 600i.

Memristive characteristics (I-V curves, $R_{off}/R_{on}$ ratio, plasticity, retention time, endurance) of the M/PPX/ITO structures were studied using a Cascade Microtech PM5 analytical probe station; the voltage pulses were supplied by a National Instruments PXIe-4140 source measure unit, programmed in LabView. All experiments were performed at room temperature.

**Results and Discussion**

**Structure.** The schematic view of the fabricated thin-film memristive structure is shown in Figure 1a. In all the electrical experiments described in this work, voltage is applied to the top electrode (Ag, Cu, Al or Ti), while the bottom electrode (ITO) is grounded. Figure 1b represents the chemical structure unit of parylene (active layer). Figures 1c and 1d show the top view microscopic image and cross sectional TEM image of the Cu/PPX/ITO memristive structure, respectively.

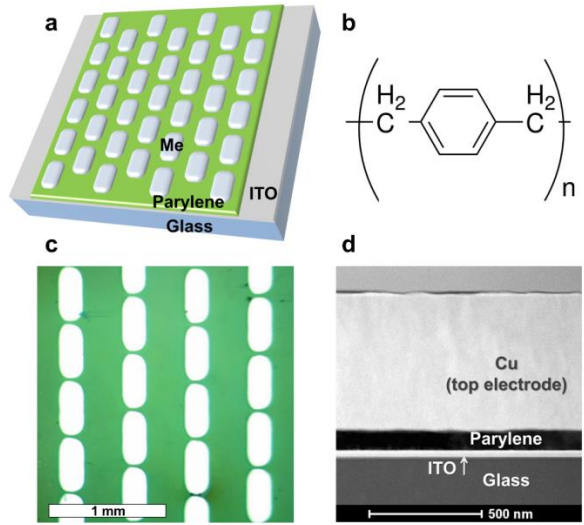

**Figure 1.** (a) Scheme of the M/PPX/ITO memristive structure. (b) Parylene N repeat unit. (c) Microscopic image of Cu/PPX/ITO memristive elements, constituting the memristive structure (only a part of the sample is shown). (d) Layers of the Cu/PPX/ITO memristive element, a TEM image.

**Memristive Characteristics.** The resistive switching characteristics of the samples could be observed from the typical cyclic I-V curves, shown in Figure 2. Compliance currents of +1 and −100 mA were set to prevent overheating and subsequent breakdown of the structures. Each cycle was carried out by applying a voltage sweep that included a rise up to a positive voltage $U_+$ sufficient to switch the M/PPX/ITO device to the low-resistance state (LRS) $(0 \rightarrow U_+ \rightarrow 0)$ and, after that, a fall down to a negative voltage $U_-$ to switch it back to the high-resistance state (HRS) $(0 \rightarrow U_- \rightarrow 0)$. The sweep rate was 2 V/s (with a step of 0.1 V). All the samples were initially in the HRS with resistances of $\sim 10^5$, $\sim 10^6$, $\sim 10^5$ and $\sim 10^5$ Ω for Ag, Cu, Al and Ti electrodes, respectively. The LRS was $\sim 1$ kΩ for all the samples.

As shown by the I-V curve in Figure 2a, the current through the Ag/PPX/ITO sample (further we will denote it as "Ag sample", the other samples will be denoted similarly) suddenly increases to the compliance level at 0.5 V during the 1st scan (number of scans is indicated next to the corresponding RS event). On the other hand, the RESET voltage was relatively high equaling

−5.5 V. It is noteworthy that the SET and RESET voltages stabilized subsequently at the values of ~ 4 V and ~ −2 V, respectively. The device exhibits stable bipolar RS behavior with $R_{off}/R_{on}$ ~ 30. The "cycle-to-cycle" reproducibility of the I-V characteristics of the Ag sample can be characterized by the SET voltage distribution (Figure 2b), representing data from ~100 I-V measurements. The maximum of the log-normal fit of this distribution is ~2 V; its mean value and standard deviation are 2.1 and 0.7 V, respectively (coefficient of variation CV = 33%).

Figure 2c demonstrates the plasticity of the Ag sample. For programming the required resistive state, a high precision (not worse than 0.5%) iterative algorithm was used, based on the application of voltage pulses with smoothly varying amplitude and duration of 100 ms [44]. As one can see from the figure, at least 8 five-minute stable resistive states (which do not intersect over time) were clearly observed between 500 Ω and 15 kΩ, what makes it possible to write at least 3 bits of information into a single memory element (the minimal number of stable resistive states was estimated in a series of retention experiments). Such stability should be enough for most short-term memory applications, but these structures also show a long-time (more than $10^3$ s) retention of low- and middle-resistance states (~ 700 and ~ 5000 Ω, see Supporting Information Figure S1). The Ag samples also show acceptable endurance (Figure 3a): more than 300 cycles of stable switching between LRS and HRS (1 and 100 kΩ, respectively). There are also prospects for increasing the number of stable resistive states since the $R_{off}/R_{on}$ value in the endurance experiment is higher than in the retention one. It should be noted that we did not observe any RS processes in Ag samples with the thickness of parylene layer ≥ 200 nm; they remained in the HRS with $R = 10^7–10^8$ Ω even when the voltage applied to the Ag electrode reached 10 V. It may indicate that the dielectric parylene coat is too thick in that case for the metal bridge to grow [1, 35].

I-V characteristics of Cu samples (Figure 2d) show a definitely better cycle-to-cycle reproducibility than in the case of the Ag top electrode. The values of $U_{set}$ are distributed (Figure 2e) with the maximum of the log-normal fit at ~ 1.3 V. The mean $U_{set}$ value of this distribution is 1.5 V, which is less than in the case of the Ag sample. The standard deviation of $U_{set}$ is 0.5 V, so the CV equals 33%. It is worth noting that the CV could be tuned by reducing the size of the memristor top electrode or by introducing additional barrier layers (e.g. graphene) [38]. On the other hand, a relatively high CV of $U_{set}$ is preferable for hardware-intrinsic security primitives (e.g. physical unclonable functions) [45]. We note that the $U_{set}$ and $U_{reset}$ parameters almost stabilize after several cycles, forming sharp-switching and symmetrical hysteretic curves. These facts are promising in terms of neuromorphic applications (because such symmetric pulses can be used for learning). Moreover, the Cu samples exhibit good plasticity; as one can see from the retention curves in Figure 2f, at least 16 stable resistive states (4 bits of information) can be programmed in memristors of this kind. Despite the fact that Figure 2 shows only short-time retention curves (5 min), the long-living states (up to 5 h) were also recorded (see Supporting Information Figure S2). The samples also demonstrate good endurance, but the gap between HRS and LRS tends to decrease slowly after ~ 600 cycles from the value of $R_{off}/R_{on} = 10^2$–$10^3$ down to ~ 10 (see Figure 3b). It may be due to the features of the measuring algorithm used: during the measurement the sample is constantly under an alternating voltage stress, which may cause its overheating and possible degradation. Going forward, we want to emphasize that in STDP measurements, where dozens of RSs to a state with preset resistance take place (according to an accurate algorithm [44]), the Cu sample exhibits good performance showing more than $10^4$ precise switchings without failure, which is the best result among the samples investigated in this work.

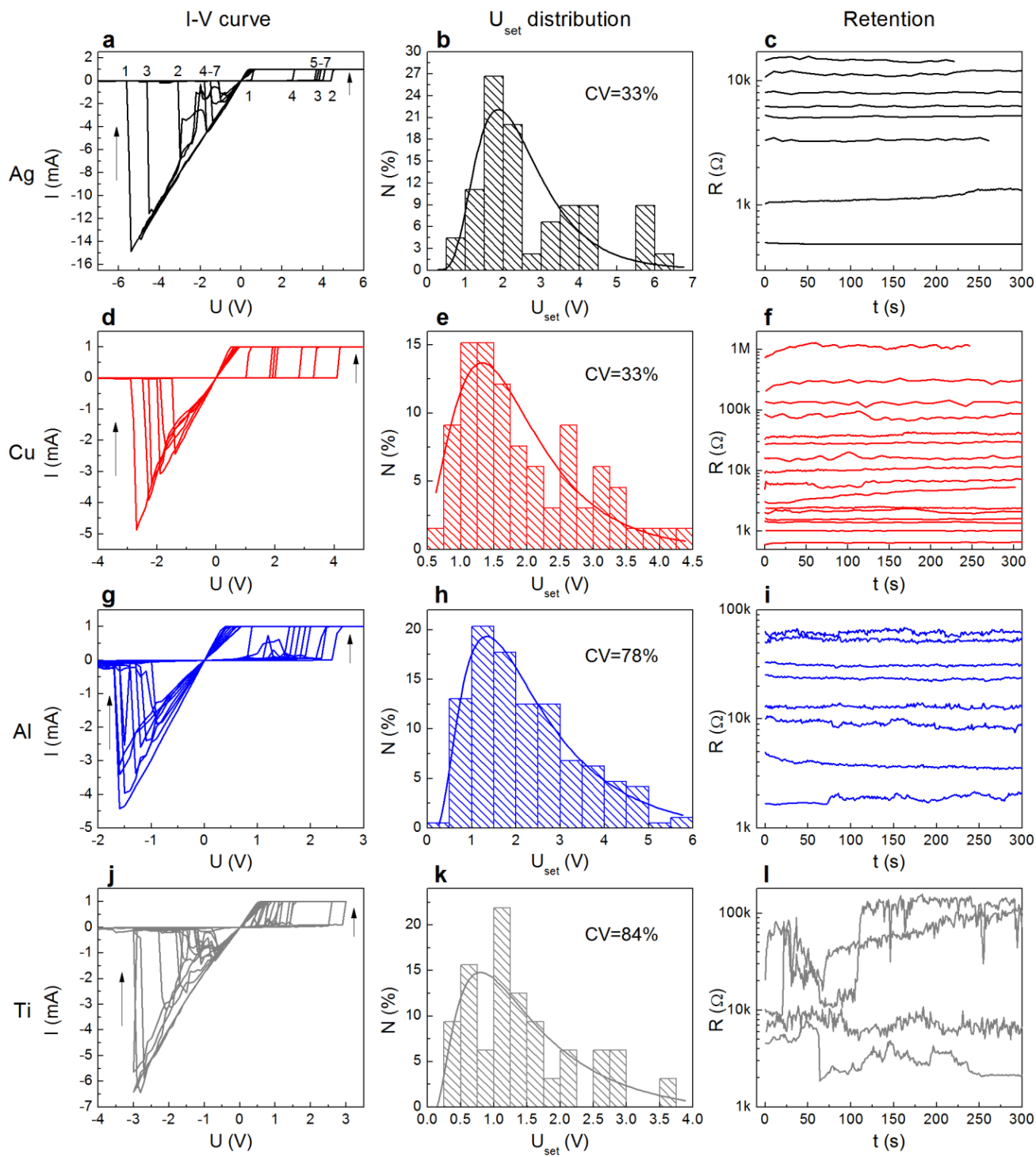

**Figure 2.** Memristive properties of (Ag, Cu, Al, Ti)/PPX/ITO structures: typical I-V curves, $U_{set}$ distributions (~ 100 cycles) and retention time ($U_{read} = 0.1$ V) characteristics.

Regarding Al samples, they have also demonstrated good cycle-to-cycle reproducibility but the $U_{set}$ distribution is wider (CV = 78%) than in the case of the Cu electrode (CV = 33%). It was estimated from the log-normal fitting (Figure 2h) that the mean value of $U_{set}$ equals here 2.7 V with the standard deviation of 2.1 V. It is too wide for neuromorphic applications. Moreover, the Al samples have worse plasticity than the Cu ones: only 8 stable resistive states were detected (Figure 2i). Their endurance is also poor: less than 100 cycles. In spite of these facts, these structures show reasonable short-time retention (see Figure 2i).

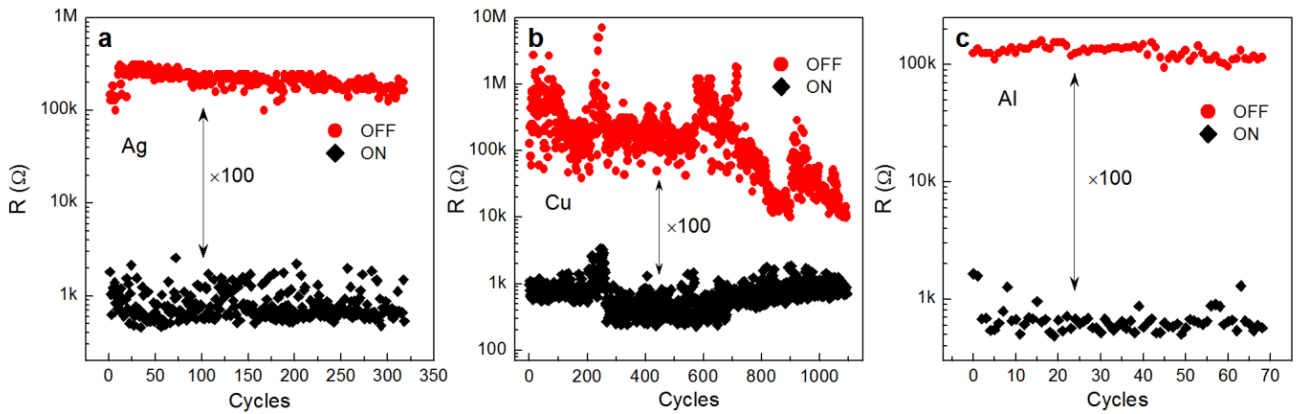

**Figure 3.** Endurance of (a) Ag/PPX/ITO, (b) Cu/PPX/ITO and (c) Al/PPX/ITO memristors. Black and red points represent low- and high-resistance states, respectively. Pulse time was 100 ms; $U_{set} = -U_{reset} = 5$ V for (a), $U_{set} = -U_{reset} = 4$ V for (b) and $U_{set} = 5$ V, $U_{reset} = -8$ V for (c).

The Ti samples, as shown in Figure 2j, do not demonstrate any of the desired features: their I-V curves are asymmetric and irreproducible, the plasticity is absent due to the poor retention, and the endurance is too noisy and insufficient (therefore it is not presented in Figure 3).

General comparison of memristive characteristics of all the samples is shown in Table 1. Note that the Cu samples are superior in all parameters, especially in plasticity. This could be explained by taking into account the higher surface activation energy of Cu compared to Ag and Al, which leads to a smaller diffusion rate [30]. Thus, the lifetime of Cu bridges is longer and, hence, the plasticity of Cu samples is better. This result is also confirmed by a recent study by Valov et al. [46], where the filament stability and retention properties of ECM devices were found to be determined by the Gibbs free energy of formation of cations.

Note that although the question of the nature of RS in M/PPX/ITO structures is extremely interesting, its detailed analysis lies beyond the scope of this study. Here, based on the experimental data and on the related literature, we only make the following conclusion: the bipolar RS behavior of the investigated samples most likely originates from the formation/destruction of metal bridges (filaments) in the dielectric parylene layer due to migration of metal cations ($Ag^+$, $Cu^+/Cu^{+2}$, $Al^{+3}$ or $Ti^{+4}$) from the top electrode toward the bottom during the SET process under strong electric field [1, 3, 30, 35, 46]. In contrast, according to the molecular dynamics simulations performed by Ielmini et al. [30], the conductive bridges can spontaneously break during the RESET switching as a result of atomic surface diffusion driven by the minimization of the system energy.

**Table 1.** Memristive Characteristics of M/PPX/ITO Samples

| Sample | max $R_{off}/R_{on}$ | $U_{set}$ ($\mu \pm \sigma$), V | Number of resistive states | Endurance, cycles |
|---|---|---|---|---|
| Ag/PPX/ITO | 100 | 2.1±0.7, CV=33% | 8 | 300 |
| Cu/PPX/ITO | 1000 | 1.5±0.5, CV=33% | 16 | 1000 |
| Al/PPX/ITO | 100 | 2.7±2.1, CV=78% | 8 | 70 |
| Ti/PPX/ITO | 50 | 1.9±1.6, CV=84% | – | – |

**Spike-Timing-Dependent Plasticity.** For the STDP implementation we chose the Cu samples because they showed the best memristive characteristics among the investigated structures. The bottom electrode (ITO) of the Cu/PPX/ITO memristive structure was assigned for the pre-synaptic input and the top electrode (Cu) was considered as the post-synaptic one. We used identical voltage pulses as pre- and post-synaptic spikes of heteropolar bi-rectangular (inset in Figure 4a) or bi-triangular (inset in Figure 4b) shape. The amplitudes of bi-rectangular and bi-triangular spikes were chosen to be 0.7 V and 0.8 V, respectively, so the spike itself could not lead to a conductivity change in the structure. On the other hand, if two spikes are summed up, the voltage drop across the memristive device could be increased up to ±1.4 and ±1.6 V, which is within the switching range of the Cu sample. The pulse half-durations were 150 and 200 ms with discretization of 50 ms. Post-synaptic pulses were applied after (before) pre-synaptic pulses with varying delay time $\Delta t$ (ranged from −500 to 500 ms with a step of 50 ms).

Conductance was measured by the application of a testing voltage of +0.1 V within 50 ms before and after the sequence of pre- and post-synaptic pulses. Generally, the device conductance $G$ is regarded as a synaptic weight, and then its change ($\Delta G$) is equal to a synaptic weight change. More specifically, a weight change corresponds to $\Delta G = G_f - G_i$, where $G_f$ and $G_i$ are the final and the initial conductance values, respectively. Thus, the determined weight change dependence on the delay time (STDP window) for two different initial states is shown in Figure 4.

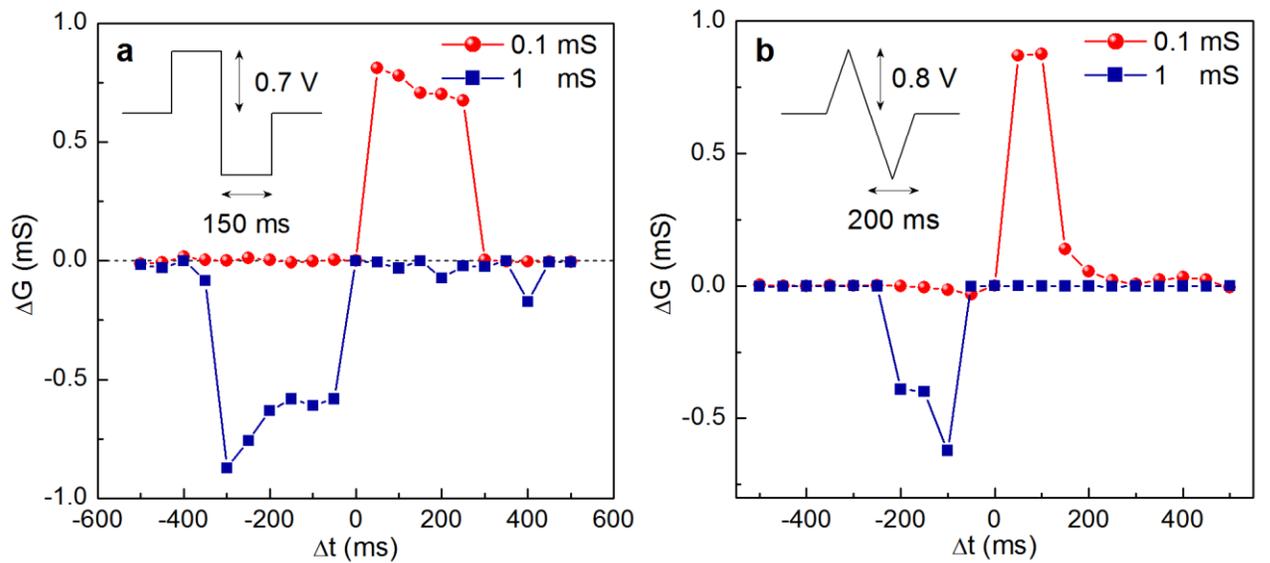

**Figure 4.** STDP window of Cu/PPX/ITO memristive structures (for various initial conductance values) obtained with heteropolar (a) bi-rectangular and (b) bi-triangular spike pulses shown in the figure insets. Post-synaptic spikes were applied after (before) pre-synaptic ones with a varying delay time $\Delta t$. Every point of the curves is a median of 10 recorded experimental values.

One can see from Figure 4 that the experimental results obey the STDP rule observed in biological systems [47]. Synaptic potentiation ($\Delta G > 0$) was observed for $\Delta t > 0$, and synaptic depression ($\Delta G < 0$) was observed for $\Delta t < 0$. Note that the result of STDP learning depends on the $G_i$ value. If a memristor state is close to LRS, then its synaptic weight would likely depress rather than potentiate (as for the 1 mS state in Figure 4), and vice versa (0.1 mS state in Figure 4). This "multiplicative" character of STDP curve can be explained by taking into account the finiteness of a conductance change in the studied memristors. Interestingly, one can observe that STDP learning with bi-triangular pulses leads to the bi-triangular-like shape of the STDP window, and the same equivalence can be traced for bi-rectangular pulses (see STDP section of Supporting Information).

It should be noted that we tested a wide range of pulse amplitudes (0.5, 0.6, 0.7, 0.8 and 0.9 V) and half-durations (50, 100, 150, 200 and 250 ms) for several resistive states of the Cu sample. These results can be found in Supporting Information (Figures S3–S8).

**Neuromorphic Application of STDP.** After successful experimental implementation of STDP for parylene-based memristive structures, we would like to take a step forward to demonstrate their utility in constructing simple neuromorphic networks capable of STDP-based learning. For this purpose, we have chosen the task of classical (also known as Pavlovian) conditioning [48, 49] and constructed a network (simulating Pavlov's dog behavior) consisting of 2 pre-synaptic neurons connected with a post-synaptic one (Figure 5a). The first pre-synaptic neuron is connected to the post-synaptic one via a resistor $R$ corresponding to an unconditioned stimulus (e.g. "food") pathway. The second pre-synaptic neuron connection is represented by a memristive element (Cu/PPX/ITO) corresponding to an initially neutral stimulus (e.g. a "bell") pathway. Each neuron was implemented in software: the pre-synaptic neurons were programmed to generate spikes of amplitude $U_{sp}$ (the spike shape was similar to that in the inset of Figure 4b), and the post-synaptic one was used as a threshold unit (generating spikes only in the case of the total input current exceeding the threshold current $I_{th}$, which is chosen to be slightly less than the ratio $U_{sp}/R$). The bottom electrode of the memristive element was connected to the output of the post-synaptic neuron. A similar electronic implementation of Pavlov's dog has been presented before [50], however there was used constant-signal learning without the use of any STDP rules. Another implementation was proposed in a network with pseudo-memcapacitive synapses with a Hebbian-like learning mechanism [51].

The learning procedure was as follows: 1) introducing a signal only down the unconditioned stimulus pathway (by this step we check the correct post-synaptic neuron activity, i.e. the dog starts "salivating", having been exposed to the sight (or smell) of "food"); 2) sending a signal only down the conditioned stimulus pathway (in this step we check whether an initially neutral stimulus becomes a conditioned one); 3) pairing the two stimuli (in this step the conditioning (learning) occurs). These three steps constituted one epoch of learning shown schematically in Figure 5b.

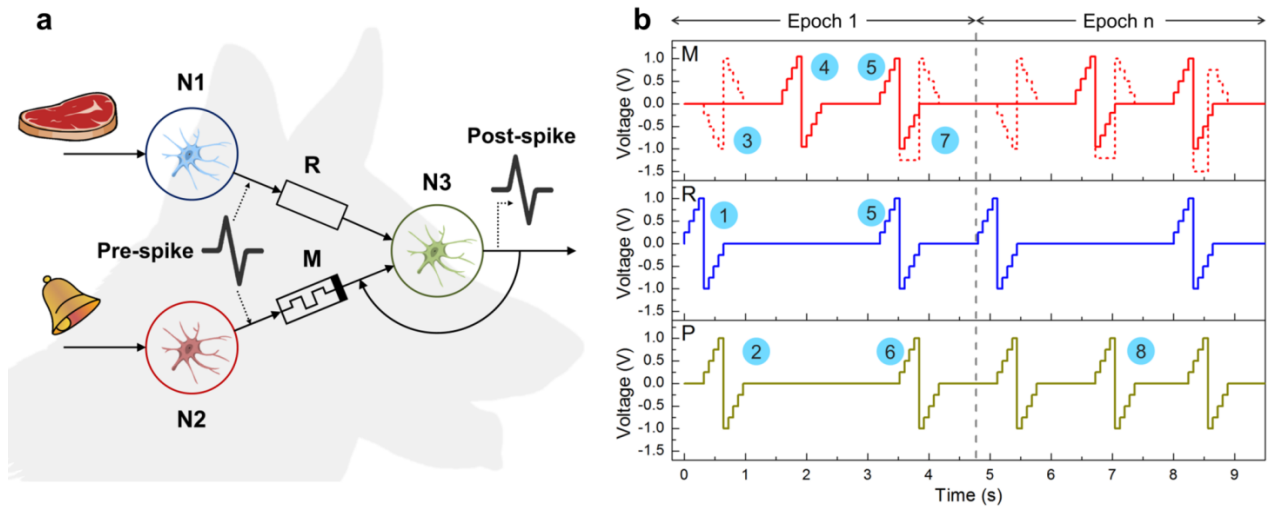

**Figure 5.** STDP-learning memristive Pavlov's dog implementation. (a) The electrical schematic diagram: N1 — the 1st pre-neuron, spiking after the "food"-related stimulus; N2 — the 2nd pre-neuron, spiking after the "bell" stimulus; N3 — the post-neuron, which spikes when the total input

current exceeds the threshold; R — a resistor with a constant resistance value $R = 2$ k$\Omega$; M — a memristive element, initially in the $R_{off} = 20$ k$\Omega$ resistive state. A post-spike is generated unconditionally after a spike comes from N1 and under the condition that the memristor current exceeds $I_{th}$ after a spike comes from N2. (b) An example of the spike pattern applied to the inputs of the scheme: 1 — the initial pulse (1st Epoch) on the resistor (R) (unconditioned stimulus), resulting in post-spike (P) 2, which in turn comes to the memristor (M) as pulse 3 (dashed) in the inverted form; 4 — the pulse on the memristor, initially without post-neuron activity; 5 — simultaneous pulses on the resistor and the memristor, which result in post-spike 6 leading to the teaching pulse 7 (dashed); 8 — a post-spike as a result of the conditioned stimulus when the training is completed (Epoch $n$, where $n$ is equal to or above the number of epochs needed for successful conditioning).

Resistance $R = 2$ k$\Omega$ (Figure 5a) was chosen to be slightly greater than the $R_{on}$ resistive state of the Cu/PPX/ITO memristive structure (~ 1 k$\Omega$) in order to provide the possibility of successful training of the network. The spike amplitudes $U_{sp}$ and durations $\Delta t_{sp}$ were identical for all neurons and were selected experimentally according to the results of the I-V and STDP measurements (see previous sections). Every experiment started with the $R_{off}$ ~ 20 k$\Omega$ state of the memristor (which was, in fact, an intermediate state, because resistance in the "true" HRS reaches 1 M$\Omega$, see Figure 2f). As is shown in Figure 5b, when the conditioned stimulus is paired with the unconditioned one (Figure 5b, spikes 5), the post-synaptic pulse (spike 6) from N3 (which starts being generated when the memristor current exceeds the current threshold $I_{th}$) sums up with the pre-synaptic pulse from N2, leading to a resistivity change in the memristor (by the dashed spike 7). If the introduction of the "bell" signal alone resulted in the post-synaptic neuron spiking (spike 8), the conditioning occurred. In our experiments the parameters were such that the threshold resistance value of the memristor that led to neuron spiking (i.e. the highest resistance value when $U_{sp}/R_{th} > I_{th}$) was $R_{th}$ ~ 5 k$\Omega$. We used spikes of heteropolar bi-triangular shape with the amplitude of 1 V and the half-duration of 80, 160 and 320 ms. It can be seen from Figure 6 that the conditioning occurred in all the cases, but the number of epochs necessary for conditioning was different depending on the pulse durations. Namely, the shorter was the pulse duration used, the larger was the number of epochs necessary for successful conditioning. This may be due to the fact that the short duration of the electric field influence (80 ms) is not enough to form a continuous metal bridge. Since the effect of local heating can be neglected here (a relatively long time passes between the pulses — about 1 s), the total pulse time remains the only parameter determining the conditioning rate (assuming constant amplitude of learning pulses). The obtained results demonstrate the associative learning ability of neuromorphic systems with parylene-based memristive elements and provide a basis for the development of autonomous circuits capable of emulating cognitive functions.

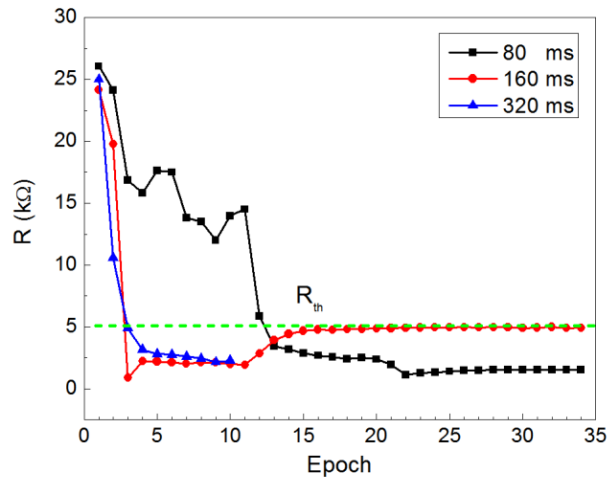

**Figure 6.** Electronic Pavlov's dog implementation: memristor resistance depending on the number of learning epochs.

**Conclusions**

In conclusion, we have successfully demonstrated that parylene-based memristive devices are capable of learning (including learning using biologically inspired STDP rules). The Metal/Parylene/ITO structures were fabricated with Ag, Cu, Al and Ti top active electrodes. These organic memristive devices (except for those with Ti top electrodes) exhibit the advantages of low switching voltage (down to 1 V), high $R_{off}/R_{on}$ ratio (up to $10^3$), long retention time ($\geq 10^4$ s) and multilevel resistance switching (at least 16 stable resistive states in the case of Cu electrodes). We have experimentally shown that parylene-based memristive elements can be trained by the STDP mechanism. The model of classical conditioning (electronic "Pavlov's dog") was implemented as a simple neuromorphic circuit using Cu/Parylene/ITO memristors. The obtained results demonstrate the associative learning ability of neuromorphic systems with parylene-based devices, which is especially valuable given that parylene is a polymer that is FDA-approved and completely safe for the human body. Thus the development of memristive systems based on it provides prospects for hardware realization of artificial neural networks for wearable and biomedical applications.

AUTHOR INFORMATION

**Corresponding Author**

*(A.A.M.) E-mail: minnekhanov@physics.msu.ru, minnekhanov_aa@nrcki.ru. Tel: +7 (926) 666-59-79

**Present Addresses**

[†]Deutsches Elektronen-Synchrotron DESY, 22607 Hamburg, Germany.

**Notes**

The authors declare no competing financial interest.

**Acknowledgments**

This work was partially supported by the Russian Science Foundation (project № 18-79-10253) in the STDP investigation part and by the Russian Foundation for Basic Research (project № 18-37-20014) in the part concerning investigation of memristive characteristics. Measurements were carried out with the equipment of the Resource Centers (NRC "Kurchatov Institute"). Authors are thankful to Prof. A. V. Sitnikov (Voronezh State Technical University) and Dr. M. L. Zanaveskin (NRC "Kurchatov Institute") for assistance with the top contact deposition and to Dr. M. Yu. Presnyakov (NRC "Kurchatov Institute") for the TEM measurements.

**Supporting Information**

Additional retention and STDP data.

**For Table of Contents Only**

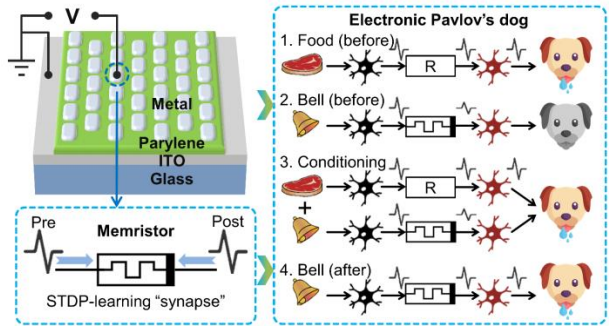

**Supporting Information**

# Parylene Based Memristive Devices with Multilevel Resistive Switching for Neuromorphic Applications

*Anton A. Minnekhanov[1,*], Andrey V. Emelyanov[1,2], Dmitry A. Lapkin[1,†], Kristina E. Nikiruy[1,2], Boris S. Shvetsov[3], Alexander A. Nesmelov[1], Vladimir V. Rylkov[1,4], Vyacheslav A. Demin[1,2], Victor V. Erokhin[1,5]*

[1]National Research Center "Kurchatov Institute", 123182 Moscow, Russia.

[2]Moscow Institute of Physics and Technology, 141700 Dolgoprudny, Moscow Region, Russia.

[3]Lomonosov Moscow State University, 119991 Moscow, Russia.

[4]Kotel'nikov Institute of Radio Engineering and Electronics RAS, 141190 Fryazino, Moscow Region, Russia.

[5]CNR-IMEM (National Research Council, Institute of Materials for Electronics and Magnetism), 43124 Parma, Italy

[†]Present address: Deutsches Elektronen-Synchrotron DESY, 22607 Hamburg, Germany.

*E-mail: minnekhanov@physics.msu.ru, minnekhanov_aa@nrcki.ru

**Long-time retention**

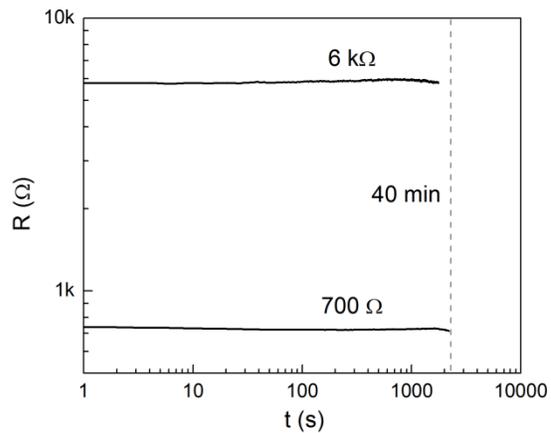

**Figure S1.** Long-time retention of Ag/PPX/ITO memristors.

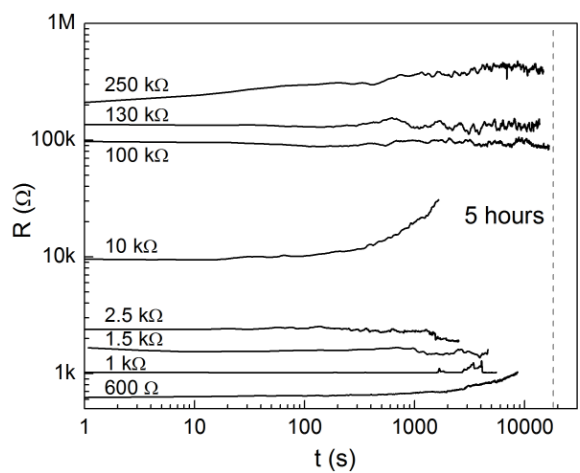

**Figure S2.** Long-time retention of Cu/PPX/ITO memristors.

**STDP**

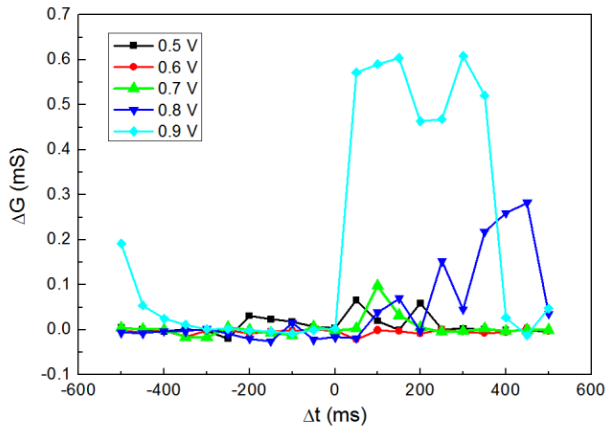

**Figure S3.** STDP window of Cu/PPX/ITO memristive structures, obtained with bi-rectangular pulses of various amplitudes. Post-synaptic STDP pulses were applied after pre-synaptic ones with a varying delay time Δ*t*. All the pulses were symmetric. Every point of the curves is a median of 10 recorded experimental values. Initial conductance was 0.1 mS, pulse half-durations were 200 ms.

As one can see, the best result is obtained when the pulses of 0.7 V are applied (the shape of the curve is smooth and it has one extremum).

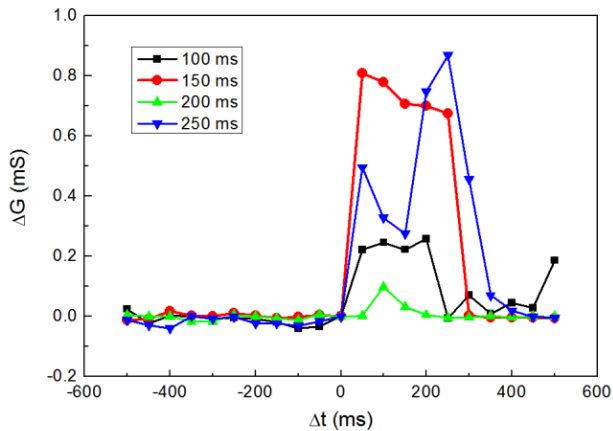

**Figure S4.** STDP window of Cu/PPX/ITO memristive structures, obtained with bi-rectangular pulses of various half-durations. Post-synaptic STDP pulses were applied after pre-synaptic ones

with a varying delay time Δ*t*. All the pulses were symmetric. Every point of the curves is a median of 10 recorded experimental values. Initial conductance was 0.1 mS, pulse amplitudes were 0.7 V.

As one can see, the best result in this case is obtained when the pulses of 150 ms are applied.

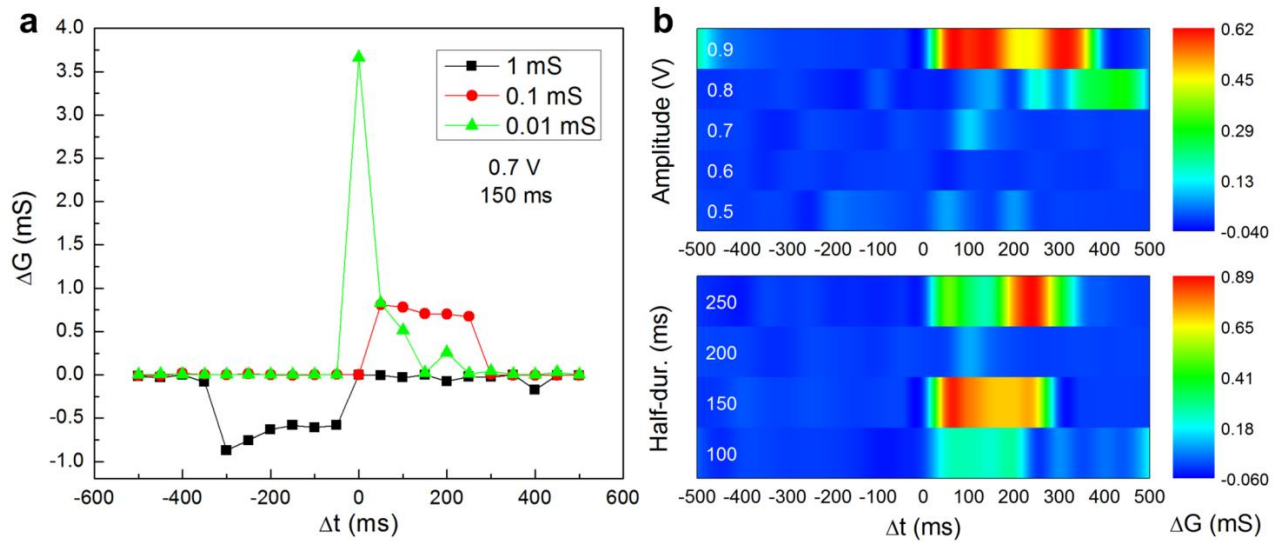

**Figure S5.** STDP window of Cu/PPX/ITO memristive structures obtained with bi-rectangular pulses. (a) STDP window for various initial conductance values. Post-synaptic STDP pulses were applied after pre-synaptic ones with a varying delay time Δ*t*. All the pulses were symmetric. Every point of the curves is a median of 10 recorded experimental values. Pulse amplitudes and half-durations were 0.7 V and 150 ms, respectively. (b) STDP window for various pulse amplitudes (top image, half-duration is 200 ms) and half-durations (bottom image, amplitude is 0.7 V). Images are obtained by color mapping of a cubic interpolation (2000 pts) of the corresponding STDP curves (Figures S4–S5).

By analogy, the bi-triangular pulse STDP was checked.

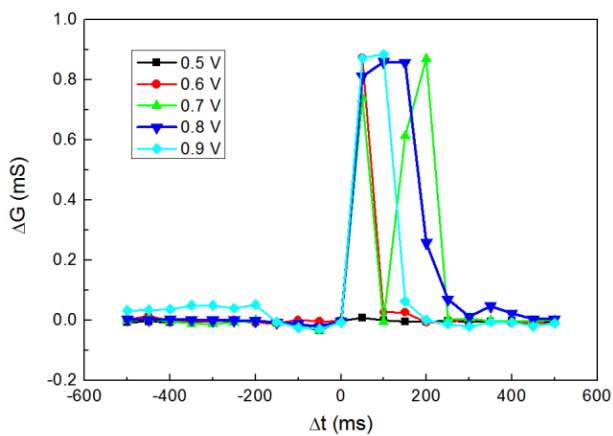

**Figure S6.** STDP window of Cu/PPX/ITO memristive structures, obtained with bi-triangular pulses of various amplitudes. Post-synaptic STDP pulses were applied after pre-synaptic ones with a varying delay time $\Delta t$. All the pulses were symmetric. Every point of the curves is a median of 10 recorded experimental values. Initial conductance was 0.1 mS, pulse half-durations were 200 ms.

Note that the 0.8 V curve is the smoothest and has one extremum, therefore we chose this pulse amplitude for the next experiments.

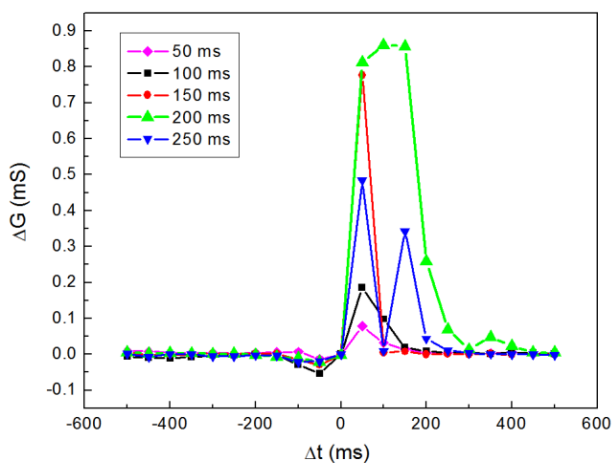

**Figure S7.** STDP window of Cu/PPX/ITO memristive structures, obtained with bi-triangular pulses of various half-durations. Post-synaptic STDP pulses were applied after pre-synaptic ones with a varying delay time Δ*t*. All the pulses were symmetric. Every point of the curves is a median of 10 recorded experimental values. Initial conductance was 0.1 mS, pulse amplitudes were 0.8 V.

The same experiment parameters (0.8 V, 200 ms) show here the best result in this case.

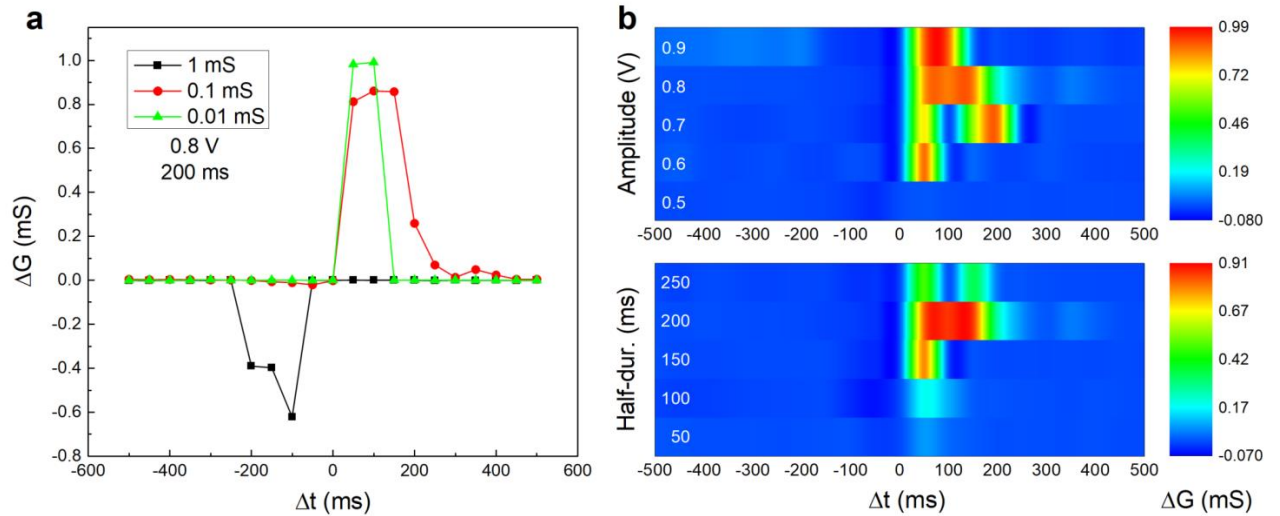

**Figure S8.** STDP window of Cu/PPX/ITO memristive structures obtained with bi-triangular pulses. (a) STDP window for various initial conductance values. Post-synaptic STDP pulses were applied after pre-synaptic ones with a varying delay time Δ*t*. All the pulses were symmetric. Every point of the curves is a median of 10 recorded experimental values. Pulse amplitudes and half-durations were 0.8 V and 200 ms, respectively. (b) STDP window for various pulse amplitudes (top image, half-duration is 200 ms) and half-durations (bottom image, amplitude is 0.8 V). Images are obtained by color mapping of a cubic interpolation (2000 pts) of the corresponding STDP curves (Figures S6–S7).